\documentclass[a4paper]{article}

\usepackage{latexsym}
\usepackage{amsmath}
\usepackage{amssymb}
\newtheorem{theorem}{Theorem}
\newtheorem{lemma}[theorem]{Lemma}

\newtheorem{prop}[theorem]{Proposition}

\begin{document}

\def\R{\mathbf{R}} 
\def\Cx{\mathbf{C}}
\def\Z{\mathbf{Z}} 
\def\x{\mathbf{x}} 
\def\del{\partial} 
\def\Lap{\bigtriangleup} 
\def\^{\wedge} 
\def\goes{\rightarrow}
\def\inv{^{-1}}

\def\reff#1{(\ref{#1})}
\def\vb#1{{\partial \over \partial #1}} 
\def\vbrow#1{{\partial/\partial #1}} 
\def\Del#1#2{{\partial #1 \over \partial #2}}
\def\Dell#1#2{{\partial^2 #1 \over \partial {#2}^2}} \def\Dif#1#2{{d
    #1 \over d #2}} \def\Lie#1{ \mathcal{L}_{#1} } \def\diag#1{{\rm
    diag}(#1)} \def\abs#1{\left | #1 \right |} \def\rcp#1{{1\over #1}}
\def\paren#1{\left( #1 \right)} \def\brace#1{\left\{ #1 \right\}}
\def\bra#1{\left[ #1 \right]} \def\angl#1{\left\langle #1
  \right\rangle} 
\def\vector#1#2#3{\paren{\begin{array}{c} #1 \\ #2 \\
      #3 \end{array}}} 
\def\svector#1#2{\paren{\begin{array}{c} #1 \\
      #2 \end{array}}}
\def\matrix#1#2#3#4#5#6#7#8#9{ \left( \begin{array}{ccc}
        #1 & #2 & #3 \\ #4 & #5 & #6 \\ #7 & #8 & #9
        \end{array}  \right) }
\def\smatrix#1#2#3#4{ \left( \begin{array}{cc} #1 & #2 \\ #3 & #4
        \end{array}  \right) }

\def\GL#1{{\rm GL}(#1)} 
\def\SL#1{{\rm SL}(#1)} 
\def\wSL#1{{\widetilde{\mathrm{SL}}}(#1)} 
\def\PSL#1{{\rm PSL}(#1)} 
\def\O#1{{\rm O}(#1)} 
\def\SO#1{{\rm SO}(#1)}
\def\IO#1{{\rm IO}(#1)} 
\def\ISO#1{{\rm ISO}(#1)} 
\def\U#1{{\rm U}(#1)} 
\def\SU#1{{\rm SU}(#1)}

\def\Teich{{Teichm\"{u}ller}} \def\Poin{{Poincar\'{e}}}

\def\Gam{\mbox{$\Gamma$}} \def\d{{d}} \def\VII#1{\mbox{VII${}_{#1}$}}
\def\VI#1{\mbox{VI${}_{#1}$}}
\def\Isom{{\mathrm{Isom}}}

\def\hh{{h}}
\def\ggg{{\rm g}} 
\def\uh#1#2{\hh^{#1#2}} 
\def\dh#1#2{\hh_{#1#2}}
\def\mh#1#2{\hh^{#1}{}_{#2}} 
\def\ug#1#2{\ggg^{#1#2}}
\def\dg#1#2{\ggg_{#1#2}} 
\def\uug#1#2{\tilde{\ggg}^{#1#2}}
\def\udg#1#2{\tilde{\ggg}_{#1#2}} 
\def\udh#1#2{\tilde{\hh}_{#1#2}}

\def\c#1{\chi_{#1}} \def\cc#1#2{\chi_{#1}{}^{#2}} \def\uc#1{\chi^{#1}}
\def\s#1{\sigma^{#1}} \def\ss#1#2{\sigma^{#1}{}_{#2}}
\def\om#1#2#3{\omega_{#1#2}{}^{#3}} \def\omd#1#2#3{\omega_{#1#2#3}}
\def\CC{C} \def\C#1#2#3{\CC^{#1}{}_{#2#3}} \def\Sig#1{\Sigma^{#1}}
\def\Sigg#1#2{\Sigma^{#1}{}_{#2}} \def\Chi#1{X_{#1}}
\def\Chii#1#2{X_{#1}{}^{#2}} \def\dug#1#2{g_{#1}{}^{#2}}
\def\dgm#1#2{\gamma_{#1#2}} \def\ugm#1#2{\gamma^{#1#2}}
\def\covsset{\mathcal{S}_2} \def\consset{\mathcal{S}^2} \def\X{X}
\def\dd#1#2{\frac{\del^2{#1}}{\del {#2}^2}}

\def\wa{\!\!\!\!&=&\!\!\!\!}  
\def\wb{\!\!\!\!&\equiv &\!\!\!\!}
\def\ws{&=} 
\def\nd{\noindent} 
\def\D{{\mathcal{D}}}

\def\proofmark{\textsc{Proof.} } 
\def\defmark{{\bf Definition}\hspace{1em}} 
\def\defsmark{{\bf Definitions}\hspace{1em}} 
\def\notemark{{\bf Note.} }
\def\remarkmark{\textsc{Remark.} } 
\def\conventionmark{{\bf Convention.}}
\def\endofproofmark{\hfill\rule{.5em}{.5em}}

\def\N{\mathcal{N}}
\def\dq#1#2{q_{#1#2}}
\def\uq#1#2{q^{#1#2}}
\def\udq#1#2{\tilde{q}_{{#1}{#2}}}
\def\bdq#1#2{{\bar q}_{#1#2}}
\def\dqstd#1#2{\tilde{q}_{#1#2}^{(0)}}
\def\bdh#1#2{{\bar h}_{#1#2}}
\def\dhstd#1#2{\tilde{h}_{#1#2}^{(0)}}
\def\barphi{{\bar\phi}}
\def\signm{\varsigma}
\def\mass{\mathrm{m}_\Phi}
\def\am{\mathsf{L}} 

\def\bgam#1{\mathbf{\gamma}_{#1}}

\def\harV{V}
\def\vEE#1#2#3{(\harV^{#1}_{#2})_{#3}}
\def\vE#1#2{\harV^{#1}_{#2}}
\def\vEEd#1#2#3{(\harV^{\prime #1}_{#2})_{#3}}
\def\vEd#1#2{\harV^{\prime #1}_{#2}}
\def\vEEdd#1#2#3{(\harV^{\prime\prime #1}_{#2})_{#3}}
\def\vEdd#1#2{\harV^{\prime\prime #1}_{#2}}

\def\b{\beta}
\def\G#1{G_{\mathrm{#1}}}

\def\AA#1{\mathcal{A}_{#1}}
\def\spec#1{{\mathrm{Spec}#1}}
\def\M{{\Gamma\backslash G}}


\def\AT{@}
\def\country{de}
\def\eml{email}

\title{Scalar fields on $\widetilde{\mathrm{SL}}(2,\mathbf{R})$ and
  $H^2\times\R$ geometric spacetimes and linear perturbations}

\author{{ Masayuki Tanimoto}
  \thanks{\eml: masayuki.tanimoto{\AT}aei.mpg.\country} \\
  \textit{Max-Planck-Institut f\"ur Gravitationsphysik} \\
  \textit{(Albert-Einstein-Institut)} \\
  \textit{Am M\"uhlenberg 1, Golm 14476, Germany}}

\maketitle

\begin{abstract}
  Using appropriate harmonics, we study the future asymptotic behavior
  of massless scalar fields on a class of cosmological vacuum
  spacetimes. The spatial manifold is assumed to be a circle bundle
  over a higher genus surface with a locally homogeneous metric. Such
  a manifold corresponds to the
  $\widetilde{\mathrm{SL}}(2,\mathbf{R})$-geometry (Bianchi VIII type)
  or the $H^2\times\R$-geometry (Bianchi III type). After a technical
  preparation including an introduction of suitable harmonics for the
  circle-fibered Bianchi VIII to separate variables, we derive systems
  of ordinary differential equations for the scalar field. We present
  future asymptotic solutions for these equations in a special case,
  and find that there is a close similarity with those on the
  circle-fibered Bianchi III spacetime.  We discuss implications of
  this similarity, especially to (gravitational) linear perturbations.
  We also point out that this similarity can be explained by the
  \textit{fiber term dominated behavior} of the two models.
\end{abstract}

\section{Introduction}

Linear perturbation analysis of Einstein's equation has fundamental
importance in general relativity. It expands the significance of an
exact solution that usually has a large symmetry by providing
additional properties of the solution, especially those about
stability or instability when the solution deviates from the symmetric
configuration. In particular, detailed properties of the perturbations
are expected to be served as useful pieces of information with which
global nonlinear analyses, like those for the global existence problem
\cite{LA,Ren} and the cosmic censorship conjecture \cite{Pen,ME} or
conformal dynamics by Einstein's flow \cite{And,FM}, can develop.

Linear perturbation analysis can be however very difficult and often
requires a formidable effort to carry out for many of the solutions we
are interested in. Among them are the spatially homogeneous solutions
\cite{EM,exct} which are not isotropic in any limit keeping the
homogeneity. Linear stability properties are not known for most of
those solutions. Recently, however, a complete perturbation analysis
for a vacuum Bianchi III solution, which belongs to this class of
anisotropic solutions, has been carried out \cite{TMY,Ta03} after an
enormous effort. In connection with this, we in this paper propose an
indirect approach that extends the Bianchi III result to another
model. More precisely, we make use of an analogy to derive
perturbation properties for a ``Bianchi VIII'' solution.

The setting is as follows. The spacetime manifold we consider is the
direct product of a spatial manifold $M$ and time $\R$, for which $M$
is a circle bundle over a higher genus surface $\Sigma_g$ with genus
$g>1$.  The background spatial metric is assumed to be locally
homogeneous. In this situation, there are two kinds of possible
locally homogeneous metrics on $M$ depending upon whether $M$ is a
nontrivial bundle or a trivial bundle. If the bundle is nontrivial the
metric becomes the Bianchi VIII type (Thurston's $\wSL{2,\R}$
\cite{Sco,Thu}), while if it is trivial it becomes the Bianchi III
type (Thurston's $H^2\times\R$).

The key fact is that the two kinds of vacuum spacetimes behave the
``same'' way in the future asymptotics in a special case; the
evolutions of the fibers become asymptotically the same, and the
evolutions of the base also become asymptotically the same. We will
call this coincidence of the two backgrounds the \textit{background
  asymptotic degeneracy}. (This ``degeneracy'' makes sense only when
considering a class of fibered spatial manifolds with the same fiber
and base manifold.) With this kind of coincidence, it is of
great interest to compare test fields like scalar fields on these
spacetimes and see if there is a substantial difference in the
asymptotic behaviors of the fields.  That is, if we find that
there is no substantial difference in those behaviors --- we will call
this coincidence the \textit{scalar field asymptotic degeneracy}, then
this may be regarded as an evidence that other kinds of fields
like the linear perturbations on the two kinds of spacetimes will also
show the same asymptotic behavior or at least a very similar behavior
to each other.

Based on this idea, in this paper we study the massless scalar field
equation on the Bianchi VIII spacetime in detail. In particular, we
separate the field equation using appropriate harmonics and find
asymptotic solutions. We then make a comparison with the corresponding
Bianchi III system, and will confirm the scalar field asymptotic
degeneracy between the two systems.  As mentioned, since the basic
properties of the perturbations of the Bianchi III solution are
already known, the comparison provides us with a conjecture about the
Bianchi VIII perturbations.

One of the unfamiliar techniques needed to carry out our analysis may
be the harmonics for the circle-fibered Bianchi VIII manifold. Since
there does not seem to exist a prior work on this subject (see however
\cite{Bon,JanEssay} for related work), we start with introducing
appropriate ones for this manifold in
\S\S.\ref{sec:action}-\ref{sec:har}.  Then, we separate the field
equation using those harmonics and find asymptotic solutions of the
reduced equations in \S\S.\ref{sec:bg}-\ref{sec:fa}. We then compare
the results with those of the Bianchi III system in
\S.\ref{sec:compari}. The final section is devoted to conclusions.

\section{$\wSL{2,\R}$  actions}
\label{sec:action}

\def\im{\mathrm{Im}} 
\def\T{\mathbf{T}} 

In this section we introduce some basic objects associated with the
Bianchi VIII manifolds as a preliminary for developing the harmonics.

Let $G=\wSL{2,\R}$ be the universal covering group of $\SL{2,\R}$, the
real special linear group of rank 2. This group is also known as the
Bianchi VIII group.  Our spatial manifold $M$ is the quotient
$\Gamma\backslash G$ of $G$ by a discrete subgroup $\Gamma\subset G$
acting freely from the left.  $M$ is assumed to be \textit{closed} (=
compact without boundary).  In such a case, the manifold $M$ is known
to become a (non-trivial) circle bundle over a higher genus hyperbolic
surface $\Sigma_g$.\footnote{More precisely, $M$ can be a
  \textit{Seifeld fiber space} over a hyperbolic orbifold \cite{Sco},
  but in this paper we assume that $M$ is one of usual fiber bundles
  as a typical case.} ($g\geq2$ is the genus.)

For given element
\begin{equation}
  g=\smatrix abcd\in \SL{2,\R},
\end{equation}
it acts on the space of direct product
$\Omega_+\times \T$ of the upper half plane
\begin{equation}
  \Omega_+\equiv \{\zeta\in \Cx |\, \im\zeta>0\}
\end{equation}
and the circle $\T=\R/2\pi\Z$,
\begin{equation}
  g:\Omega_+\times\T\goes \Omega_+\times\T,
\end{equation}
as
\begin{equation}
  g\cdot(\zeta,z)= (\varphi_g(\zeta), z-\vartheta_g(\zeta)),
\end{equation}
where $(\zeta,z)=(x+iy,z)\in\Omega_+\times\T$,
and
\begin{equation}
  \varphi_g(\zeta)\equiv \frac{a\zeta+b}{c\zeta+d},\quad
  \vartheta_g(\zeta)\equiv 2\,\mathrm{arg}(c\zeta+d).
\end{equation}
In fact, it is a direct computation to confirm the homomorphism
\begin{equation}
  (gg')\cdot (\zeta,z)= g\cdot (g'\cdot(\zeta,z)),
\end{equation}
where $gg'$ is understood as the usual matrix product.  Since this
natural action is simply transitive, i.e., for arbitrary $\x,\x'\in
\Omega_+\times\T$ there exists unique $g\in\SL{2,\R}$ such that
$g\cdot\x=\x'$, we can identify the space $\Omega_+\times\T$ with the
group $\SL{2,\R}$ by associating $g\in \SL{2,\R}$ with its action on
the origin $\x_0=(0+1 i,0)$; $g\simeq g\cdot
\x_0$. If we take the universal cover of $\Omega_+\times\T$, it
apparently becomes $\Omega_+\times\R$, which can be also identified
with the space $\R^3_+\equiv \{(x,y,z)\in\R^3|y>0\}$, therefore
\begin{equation}
  G=\wSL{2,\R}\simeq\Omega_+\times\R\simeq\R^3_+.
\end{equation}
This identification is understood throughout this paper. The following
explicit formula for the correspondence $G\simeq\R_+$ is sometimes
useful;
\begin{equation}
  \label{eq:GR+}
  \rcp{\sqrt y}
  \smatrix{y\cos\frac z2-x\sin\frac z2}{y\sin\frac z2-x\cos\frac z2}
  {-\sin\frac z2}{\cos\frac z2}
  \simeq (x,y,z).
\end{equation}

To find generators of $G$ acting on $\R^3_+$,
it is convenient to introduce the following one-parameter subgroups of
$\SL{2,\R}$;
\begin{equation}
  n_s= \smatrix{1}{s}{0}{1},\;
  a_t= \smatrix{e^{t/2}}00{e^{-t/2}},\;
  u_\theta=
  \smatrix{\cos \theta/2}{\sin \theta/2}{-\sin \theta/2}{\cos
  \theta/2},
\end{equation}
where $\theta\in [0,4\pi)$, $t\in\R$, and $s\in\R$.
Then, any element $g\in\SL{2,\R}$ is uniquely expressed in the form
(\cite{Su}, Proposition 1.3, Chap V)
\begin{equation}
  g=u_\theta a_t n_s.
\end{equation}
This decomposition is known as the Iwasawa decomposition.
Let $\xi_1$, $\xi_2$, and $\xi_3$ be the generators of $n_s$, $a_t$,
and $u_\theta$, respectively; so, e.g., for $\x\in\R^3_+$,
\begin{equation}
  \xi_1f(\x)=\frac{d}{ds}f(n_s\cdot\x)\big|_{s=0}=\vb xf(\x).
\end{equation}
Similarly, we can compute the others, and find
\begin{equation}
  \begin{split}
    \xi_1 &= \vb x, \\
    \xi_2 &= x\vb x+y \vb y, \\
    \xi_3 &= \rcp{2}\paren{{x^2}-{y^2}+1}\vb x
    +xy\vb y+y\vb z.
  \end{split}
\end{equation}
These generators satisfy the following algebra;
\begin{equation}
  [\xi_1,\xi_2]=\xi_1,\quad
  [\xi_2,\xi_3]=-\xi_1+\xi_3,\quad
  [\xi_3,\xi_1]=-\xi_2.
\end{equation}
Although this is not of the canonical form that is usually assumed as
the algebra of Bianchi VIII, it is easy to see that the linear
combinations
\begin{equation}
  \xi'_1= \rcp{\sqrt2}(-\frac32\xi_1+\xi_3),\quad
  \xi'_2= \xi_2,\quad
  \xi'_3= \rcp{\sqrt2}(\frac12\xi_1+\xi_3),
\end{equation}
satisfy
\begin{equation}
    [\xi'_I,\xi'_J] = \C KIJ\xi'_K
\end{equation}
with the structure constants $\C KIJ=-\C KJI$ for which the
nonzero independent components are given by
\begin{equation}
  \label{eq:SC}
  \C 123=\C 231=1,\; \C 312=-1.
\end{equation}
This coincides with the canonical choice of structure constants of
Bianchi VIII (e.g., \cite{EM}).

The generators $\xi_I$ above are regarded as generating the
\textit{left action} of $G$ on $G\simeq\R^3_+$. Another important kind
of objects are generators of the \textit{right action}, which we
denote as $\c I$. They are defined as differential operators such
that they commute with all the left generators $\xi_I$
\begin{equation}
  \label{eq:comcxi}
  [\c I,\xi_J]=0, \quad (I,J=1\sim 3),
\end{equation}
and possess the same structure constants \reff{eq:SC}. They are given
by
\begin{equation}
  \label{eq8:invv}
  \begin{split}
  \chi_1 &= y\paren{\sin z\vb x-\cos z\vb y}-\sin z \vb z,\\
  \chi_2 &= y\paren{\cos z\vb x+\sin z\vb y}-\cos z \vb z,\\
  \chi_3 &= \vb z.
  \end{split}
\end{equation}
The commutativity \reff{eq:comcxi} means that $\c I$ are
invariant under the left action of $G$ as seen from the vanishing of
the Lie derivatives, $\Lie{\xi_I}\c J=0$. We therefore call $\c I$ the
\textit{invariant vectors}.
The dual one-forms of these vectors, the \textit{invariant
  one-forms}, can be expressed as
\begin{equation}
  \label{eq8:inv1}
  \begin{split}
  \s1 &= \rcp y(\sin z dx-\cos z dy), \\
  \s2 &= \rcp y(\cos z dx+\sin z dy), \\
  \s3 &= \frac{dx}y+dz.
  \end{split}
\end{equation}
These one-forms are invariant in the sense $\Lie{\xi_I}\s J=0$, and
satisfy the following Maurer-Cartan relation
\begin{equation}
  d\s I = -\rcp2 \C IJK \s J\^ \s K,
\end{equation}
with respect to the same structure constants \reff{eq:SC}. Now, a
homogeneous metric on $G$ can be expressed as $q=\dq IJ\s I\s J$ with
$\dq IJ$ being constants, since we apparently have $\Lie{\xi_I}q=0$
for this metric. The generators $\xi_I$ of the left actions are
\textit{Killing vectors} for these homogeneous metrics. When the
metric is of the form $q=q_1((\s1)^2+(\s2)^2)+q_3(\s3)^2$, where $q_1$
and $q_3$ are constants, this metric possesses a fourth Killing vector
\begin{equation}
  \xi_4=\vb z,
\end{equation}
and the metric is said to be \textit{locally rotationally symmetric
  (LRS)}. Note that this additional Killing vector for an LRS metric
coincides with one of the invariant vectors, $\c3$;
\begin{equation}
  \label{eq:xi4=c3}
 \xi_4=\c3.
\end{equation}

An importance of the invariant vectors $\c I$ is that they generate
the \textit{regular representation} $(T,L^2(\Gamma\backslash G))$ of
$G$ on the Hilbert space $L^2(\Gamma\backslash G)$ of all
square-integrable functions on $\Gamma\backslash G$. By regular
representation, we mean the ``right regular
representation.''\footnote{For given $g\in G$, we can define the map
  $T_g: f(\x)\goes f(\x\cdot g)$, where $\x\in G\simeq \R^3_+$. The
  right action $\x\cdot g$ is naturally defined viewing $\x$ as the
  corresponding element in $G$.  It is easy to confirm that the map
  $T: g\goes T_g$ is a homomorphism; $T_gT_{g'}=T_{gg'}$.  This
  homomorphism $(T,L^2(G))$ is called the (right) \textit{regular
    representation} of $G$ on $L^2(G)$.\cite{Su}} It is apparent that
$\c I$ generate the regular representation on $L^2(G)$, since $\c I$
are generators of the right action (of $G$) on $G$.  Note that the
invariant vectors $\c I$ are naturally well defined on the quotient
$\Gamma\backslash G$, as well, since $\c I$ are by definition
invariant under the action of $\Gamma\subset G$. (Therefore for the
covering map $\pi: G\goes \Gamma\backslash G$, the induced vectors
$\pi^*\c I$ on $\Gamma\backslash G$ are well defined, which however we
simply denote as $\c I$.) Moreover, since the right and left actions
commute each other the right action of $G$ on the left quotient
$\Gamma\backslash G$ is also well defined. Therefore the regular
representation $(T,L^2(\Gamma\backslash G))$ is well defined and $\c
I$ are its generators.

With the aid of Eq.\reff{eq:GR+}, it is easy to explicitly compute the
right action $\x\cdot g$ of $g\in \SL{2,\R}$ on $\x\in\R^3_+$.
(Compute the usual matrix product of the left hand side of
Eq.\reff{eq:GR+} and $g$, and then read off the components in $\R_+$,
using the correspondence \reff{eq:GR+}.) The most important action is
that of the compact subgroup $u_\theta$, which is found to cause the
translations $(x,y,z)\goes (x,y,z+\theta)$. Its generator is therefore
found to coincide with $\c3$, which we call the \textit{fiber
  generator}.

Eq.\reff{eq:xi4=c3} is saying that if the metric is LRS, the fiber
generator $\c3$ is also a Killing vector. Therefore an LRS metric of
Bianchi VIII type is $\U1$-\textit{symmetric} (as well as being
locally homogeneous), since there exists isometries along the circle
($\simeq\U1$) fibers. Beware that non-LRS Bianchi VIII metrics are
\textit{not} $\U1$-symmetric in any sense despite that they are
circle-fibered and locally homogeneous. This is apparent from the fact
that none of the generic Killing vectors $\xi_I$ ($I=1\sim3$) are well
defined on the quotient manifold $\M$, since $\xi_I$ are not
commutative with each other. (Remember that the action of $\Gamma$ is
generated by $\xi_I$. This noncommutativity therefore implies the
noncommutativity between $\Gamma$ and $\xi_I$, which in turn implies
$\xi_I$ are not well defined on $\M$.)

The regular representation $(T,L^2(\Gamma\backslash G))$ is, as shown
in the next section, unitary, but not irreducible. As well known
(e.g., \cite{Su}), harmonics are basis vectors of \textit{irreducible
  components} of a regular representation.  We are hence interested in
irreducible components in $(T,L^2(\Gamma\backslash G))$, which are the
subject of the next section.

\section{The harmonics}
\label{sec:har}

The most important entity to consider an irreducible unitary
representation is the Casimir operator, which is an operator which
commutes with all the generators of the representation, since from
Schur's lemma, such an operator must be a constant when acting on an
irreducible space. For the group $G$, it is given by
\begin{equation}
  \square\equiv (\c1)^2+(\c2)^2-(\c3)^2.
\end{equation}
In fact, it is a direct computation to check the condition
$[\square,\c I]=0$ for all $I=1\sim 3$. In particular, it commutes
with the fiber generator $\c3$;
\begin{equation}
  [\square,\c3]=0.
\end{equation}
It is therefore possible to simultaneously diagonalize these two
operators;
\begin{equation}
  \label{eq:eigeneqns}
\begin{split}
  \square\phi_{m,\Lambda} &=-\Lambda \phi_{m,\Lambda}, \\
  \c3\phi_{m,\Lambda} &= i m \phi_{m,\Lambda}.
\end{split}
\end{equation}
We call $\Lambda$ the \textit{Casimir eigenvalue}, $m$ the
\textit{fiber eigenvalue}.

We assume that the closed manifold $\M$ coincides with a
compactification of $\SL{2,\R}$ (rather than $G=\wSL{2,\R}$). This is
equivalent to assuming that the discrete subgroup $\Gamma$ has the
subgroup $2\pi\Z$ generated by the action $(x,y,z)\goes (x,y,z+2\pi)$.
Any two points $(x,y,z)$ and $(x,y,z+2\pi)$ should therefore be
identified, which forces
\begin{equation}
  \label{eq:minZ}
  m\in\Z.
\end{equation}
If we considered a $p$-fold covering of the manifold, we would instead
have $m\in\Z /p\equiv\{n/p|n\in\Z\}$. For simplicity, however, we do
not consider this generalization in this paper.

We assume that each $\phi_{m,\Lambda}$ is normalized to unity;
\begin{equation}
  \label{eq:phinormalization}
  ||\phi_{m,\Lambda}||^2 \equiv
  (\phi_{m,\Lambda},\phi_{m,\Lambda})=1,
\end{equation}
with respect to the inner product $(f_1,f_2)$ in $L^2(\Gamma\backslash
G)$ defined by
\begin{equation}
  (f_1,f_2)\equiv \int_{\Gamma\backslash G}f_1f_2^*d\mu_0.
\end{equation}
Here, $f^*$ is the complex conjugate of $f$. The measure defined by
\begin{equation}
 d\mu_0\equiv \s1\^\s2\^\s3 
\end{equation}
is the natural left-invariant measure of $G$, therefore the above
integral on the quotient $\M$ is well defined. Since the measure is
also right-invariant (i.e., it is ``unimodular''), the above
inner product is invariant under the right action of $G$, i.e., for
$g\in G$,
\begin{equation}
  \int_{\Gamma\backslash G}f_1(\x\cdot g)f_2^*(\x\cdot g)d\mu_0=
  \int_{\Gamma\backslash G}f_1(\x)f_2^*(\x)d\mu_0.
\end{equation}
This implies that the regular representation $(T,L^2(\Gamma\backslash
G))$ is certainly \textit{unitary}.

Our main purpose here is to find the harmonics belonging to given
$\Lambda$, i.e., to find all the basis functions $
\{\phi_{m',\Lambda}\}_{m'}$ of an irreducible space labeled by
$\Lambda$. (The number of the irreducible spaces belonging to the same
$\Lambda$, the ``multiplicity,'' can be more than one. So, $\Lambda$
is not a complete label to specify the irreducible space. More about
the multiplicity will be mentioned at the end of this section.) This
task of finding the harmonics can be done starting from one function
$\phi_{m,\Lambda}$ satisfying Eqs.\reff{eq:eigeneqns}, as shown below.

First, let us define
\begin{equation}
  \label{eq8:As}
  \AA1\equiv \rcp{\sqrt2} (\chi_1-i\chi_2),\quad
  \AA2\equiv \rcp{\sqrt2} (\chi_1+i\chi_2),\quad
  \AA3\equiv -i\chi_3.
\end{equation}
We have the following properties about adjointness:
\begin{lemma}
  In the Hilbert space $L^2(\Gamma\backslash G)$, it holds
  \begin{equation}
    \label{eq:adjointAs}
  \AA1^\dagger=-\AA2,  \quad \AA3^\dagger=\AA3.  
  \end{equation}
\end{lemma}

\proofmark
We first show that the invariant operators $\c I$ ($I=1\sim 3$) are
anti-selfadjoint; $\c I^\dagger=-\c I$. Actually, if it is the case, 
$\AA1^\dagger= \rcp{\sqrt2} (\chi_1-i\chi_2)^\dagger=\rcp{\sqrt2}
(-\chi_1-i\chi_2)= -\AA2$, and $\AA3^\dagger=
-(i\chi_3)^\dagger=-i\c3=\AA3$. To show the anti-selfadjointness of
$\c I$, note the identity
\begin{equation}
  (\c If_1,f_2)=\mathcal{I}_I-(f_1,\c If_2),
\end{equation}
where we want to show the integral
\begin{equation}
  \mathcal{I}_I\equiv\int_\M \c I(f_1f_2^*)d\mu_0
\end{equation}
vanishes for all $I=1\sim 3$. Actually, we have
\begin{equation}
  \mathcal{I}_I=\int_\M \c I(f_1f_2^*)\s1\^\s2\^\s3
  =\rcp2\int_\M d(f_1f_2^*\epsilon_{IJK}\s J\^\s K),
\end{equation}
which is confirmed using the identity
\begin{equation}
 df=(\c1f)\s1+(\c2f)\s2+(\c3f)\s3,
\end{equation}
and the relation 
\begin{equation}
  d(\epsilon_{IJK}\s J\^\s K)=0,
\end{equation}
where $\epsilon_{IJK}$ is the unit skew symmetric symbol;
$\epsilon_{IJK}=\epsilon_{[IJK]}$, $\epsilon_{123}=1$. Then, from
Stokes theorem the quantity $\mathcal{I}_I$ does vanish, since the
manifold $\M$ is assumed to be closed.  \endofproofmark

\medskip

Next, note that the commutation relations among $\AA I$ become
\begin{equation}
  \label{eq8:comA}
  [\AA3,\AA1]=\AA1,\quad [\AA3,\AA2]=-\AA2,\quad [\AA1,\AA2]=\AA3.
\end{equation}
This in particular shows that operators $\AA1$ and $\AA2$ are,
respectively, the \textit{raising and lowering operator}.  In fact,
since
\begin{equation}
  \label{eq8:up}
  \AA3 \AA1\phi_m=([\AA3,\AA1]+\AA1\AA3)\phi_m=(\AA1+\AA1\AA3)\phi_m=
  (1+m)\AA1\phi_m,
\end{equation}
$\AA1\phi_m$ is an eigenfunction for $m'=m+1$;
$\AA1\phi_m\propto \phi_{m+1}$. Similarly,
$\AA2\phi_m\propto \phi_{m-1}$.
We can therefore assume that
\begin{equation}
  \label{eq:asumpab}
\begin{split}
  \AA1\phi_{m} &= a_m\phi_{m+1}, \\
  \AA2\phi_{m} &= b_m\phi_{m-1}, \\
  \AA3\phi_{m} &= m\phi_{m},
\end{split}
\end{equation}
for appropriate coefficients $a_m$ and $b_m$.

On the other hand, one can easily check the identity
\begin{equation}
  \AA1\AA2=\frac{1}{2}(\square-(\AA3)^2+\AA3).
\end{equation}
Together with the assumption \reff{eq:asumpab} we find
\begin{equation}
  \label{eq:a_{m-1}b_{m}}
  a_{m-1}b_{m} = \frac{-1}{2}\paren{(m-\rcp2)^2+\Lambda-\rcp4}.
\end{equation}

From Eq.\reff{eq:adjointAs},
we have
\begin{equation}
  \begin{split}
    (\AA1\AA2\phi_{m,\Lambda},\phi_{m,\Lambda})&=
    (\AA2\phi_{m,\Lambda},-\AA2\phi_{m,\Lambda}) \\
    &= -||\AA2\phi_{m,\Lambda}||^2 \\
    &= -|b_m|^2,
  \end{split}
\end{equation}
where we have used the normalization \reff{eq:phinormalization}.
On the other hand,
\begin{equation}
  \begin{split}
    (\AA1\AA2\phi_{m,\Lambda},\phi_{m,\Lambda})&=
    a_{m-1}b_m(\phi_{m,\Lambda},\phi_{m,\Lambda}) \\
    &= a_{m-1}b_m.
  \end{split}
\end{equation}
Therefore, using Eq.\reff{eq:a_{m-1}b_{m}},
\begin{equation}
  \label{eq:b_m^2}
  |b_m|^2=-a_{m-1}b_m=\rcp2\paren{(m-\rcp2)^2+\Lambda-\rcp4}.
\end{equation}
In particular, since $|b_m|^2\geq0$, for any possible
$m$ and $\Lambda$, it must hold that
\begin{equation}
  (m-\rcp2)^2+\Lambda-\rcp4\geq0.
\end{equation}
Since an $m=0$ mode is always contained in any irreducible
components,\footnote{This can be seen from Eq.\reff{eq:minZ} and the
  fact that $\AA1$ and $\AA2$ raises or lower the fiber eigenvalue by
  $1$. That is, $\spec{\AA3}$, the spectra of $\AA3$ for given
  $\Lambda$, always coincides with the whole range of possible $m$;
  $\spec{\AA3}=\Z$. ($\Lambda=0$ is an exceptional case, but $m=0$
  exists in this case, too.) Remark however that this would not be the
  case if $m\in\Z/p$ ($p>1$), which was as mentioned possible if
  considering a covering.} from the above condition $\Lambda$ should
be nonnegative
\begin{equation}
  \Lambda\geq0.
\end{equation}
Also,
\begin{equation}
  \label{b_m^*}
  b_m^*=-a_{m-1}.
\end{equation}

We can determine the coefficients $a_m$ and $b_m$ from
Eqs.\reff{eq:b_m^2} and \reff{b_m^*}. For convenience, let us define
\begin{equation}
  s\equiv\pm\sqrt{|\Lambda-\rcp4|}.
\end{equation}

\textbf{(i) Case $\Lambda\geq \rcp4$}

In this case, it is convenient to choose
\begin{equation}
  \label{eq:casei}
  \begin{split}
  a_m &= \frac{1}{\sqrt2}(|m+\rcp2|+is), \\
  b_m &= \frac{-1}{\sqrt2}(|m-\rcp2|-is).
  \end{split}
\end{equation}

\textbf{(ii) Case $0\leq \Lambda< \rcp4$}

In this case, we can choose
\begin{equation}
  \label{eq:caseii}
  \begin{split}
    a_m &= \frac{1}{\sqrt2}\paren{(m+\rcp2)^2-s^2}^{1/2}, \\
    b_m &= \frac{-1}{\sqrt2}\paren{(m-\rcp2)^2-s^2}^{1/2}.
  \end{split}
\end{equation}
When $\Lambda=0$ ($s^2=1/4$), from Eqs.\reff{eq:caseii} we have
$a_0=b_0=0$, reflecting the fact that this representation is trivial
(see below) and the representation space is spanned by only one
constant function $\phi_{0,0}$.  The other representations are all
infinite dimensional.

The differential representations \reff{eq:asumpab} with the
coefficients \reff{eq:casei} or \reff{eq:caseii} above provide the
heart of the relations used to separate the field equations.

In the classification of irreducible unitary representations of
$\SL{2,\R}$, there are five kinds of series (e.g., \cite{Su,Tay}). The
case (i) above belongs to the class called the \textit{first principal
  series}, while the case (ii) the \textit{complementary series}. The
other series may not generically occur from the representation we are
considering, for which the fact \reff{eq:minZ} holds.

Note that the recursion relations \reff{eq:asumpab} imply that
$\phi_{m,\Lambda}$ can be constructed by successively applying
appropriate operators on $\phi_{0,\Lambda}$. Explicitly, the relation
is given by
\begin{equation}
  \label{eq:phimphi0}
  \phi_{m,\Lambda}=
  \begin{cases}
    (a_{m-1}\inv \AA1)(a_{m-2}\inv \AA1)\cdots 
    (a_{0}\inv \AA1)\phi_{0,\Lambda}
    & (m>0) \\
    (b_{m+1}\inv \AA2)(b_{m+2}\inv \AA2)\cdots 
    (b_{0}\inv \AA2)\phi_{0,\Lambda}
    & (m<0).
  \end{cases}
\end{equation}

The function $\phi_{0,\Lambda}$ is a $z$-independent function, since
from definition $\c3\phi_{0,\Lambda}=\del\phi_{0,\Lambda}/\del z=0$.
Moreover, let us observe the fact that when acting on a
$z$-independent function $\hat f(x,y)$, the Casimir operator $\square$
degenerates to a two-dimensional Laplacian $\hat\Lap_0$. (We use hat {
  $\hat{}$ } for quantities on a surface or $z$-independent
functions.) In fact, since we can compute, from the expression
\reff{eq8:invv} of $\c I$,
\begin{equation}
  \label{eq:c12c22}
  (\c1)^2+(\c2)^2=y^2(\frac{\del^2}{\del x^2}+\frac{\del^2}{\del
  y^2})-2y\frac{\del^2}{\del x \del z}+\frac{\del^2}{\del z^2},
\end{equation}
we have
\begin{equation}
  \begin{split}
  \square\hat f &= ((\c1)^2+(\c2)^2-(\c3)^2)\hat f \\
  &= ((\c1)^2+(\c2)^2)\hat f \\
  &= \hat\Lap_0 \hat f,
\end{split}
\end{equation}
where
\begin{equation}
  \label{eq:hatLap0}
  \hat\Lap_0=y^2(\frac{\del^2}{\del x^2}+\frac{\del^2}{\del y^2}).
\end{equation}
The operator $\hat\Lap_0$ does coincide with the Laplacian, associated
with the standard two-dimensional hyperbolic metric
$h\equiv(dx^2+dy^2)/y^2=(\s1)^2+(\s2)^2$ on $\Sigma_g$.

Therefore we can summarize the construction of the harmonics on $\M$
as follows; we first consider a solution of the two-dimensional
eigenvalue equation
\begin{equation}
  \label{eq:eigeneqnonSigmag}
  \hat\Lap_0\hat\phi_{\Lambda}=-\Lambda\hat\phi_{\Lambda}
\end{equation}
on the hyperbolic surface $(\Sigma_g,h)$, and then apply the formula
\reff{eq:phimphi0} assuming $\phi_{0,\Lambda}=\hat\phi_{\Lambda}$ to
obtain an irreducible set of harmonics. Repeat this procedure for all
independent solutions for given $\Lambda$. Repeat this for all
possible values of $\Lambda$, then we obtain a complete set of
harmonics.

This construction implies a simple but remarkable fact \cite{Tay}; the
\textit{multiplicity} in an irreducible representation specified by
$\Lambda$, occurring in $(T,L^2(\M))$, is directly determined by the
multiplicity in the corresponding eigenstate of the two-dimensional
Laplacian $\hat\Lap_0$.

The spectrum of (minus) the Laplacian $-\hat\Lap_0$ on a hyperbolic
closed surface has been one of the major subjects in Riemannian
geometry \cite{Bu}. In particular, those in the range $(0,1/4)$ are
called \textit{small eigenvalues}, and in what conditions they appear
has been one of the central issues. We do not discuss this further
here, but remark that small eigenvalues appear only when $\Sigma_g$
takes a particular ``shape'' (specified by some ``Teichm\"uler
parameters''), in particular they do not necessary exist for an
arbitrarily ``shape'' of $\Sigma_g$.  The eigenvalue $\Lambda=0$
corresponds to the trivial case, so the multiplicity (in the
irreducible representation for $\Lambda=0$) is always one.

\section{The background solutions}
\label{sec:bg}

We assume that the background spacetime metric is (locally) of the
form
\begin{equation}
  \label{eq:bmetric}
  \ggg= -N^2(t)dt^2+q_1(t)(\s1)^2+q_2(t)(\s2)^2+q_3(t)(\s3)^2.
\end{equation}
No general vacuum solution for this metric is known. However, see
\cite{Hans} for the future asymptotic behavior.

On the other hand, the general vacuum solution for the LRS metric,
i.e., the case $q_1(t)=q_2(t)$, is
known. It is a special case of the so-called NUT solutions
\cite{NUT}. The Bianchi VIII NUT solution is given by
\begin{equation}
  \label{eq:LRSsol}
  N(t)^2= U(t)\inv,\quad
  q_1(t)= q_2(t)=t^2+l^2,\quad
  q_3(t)= 4l^2 U(t),
\end{equation}
where
\begin{equation}
  U(t)\equiv \frac{t^2-l^2+2\mu t}{t^2+l^2},
\end{equation}
with $l>0$ and $\mu$ being constant parameters of solution. (This form
of solution is obtained as a special case of the metric derived in
\cite{CD}.) Note that the conformal metric $h= (\s1)^2+(\s2)^2$ is the
hyperbolic metric in the upper half plane; $h=(dx^2+dy^2)/y^2$, thus
the conformal factor $q_1(=q_2)$ is the scale factor for the base
surface, while $q_3$ is the one for the circle fibers. To describe the
future asymptotic behavior of these scale factors, let us introduce a
proper time $\tau$ for this LRS solution by
\begin{equation}
  \label{eq:LRSpropertime}
  \begin{split}
  \tau &= \int Ndt= \int U^{-1/2}dt
  = \int \paren{1-\frac{\mu}{t}+O(\rcp{t^2})} dt \\
  &= t-\mu\log t+O(\rcp t).
  \end{split}
\end{equation}
This implies
\begin{equation}
  \label{eq:LRSpropertimeinv}
  \begin{split}
    t &= \tau+\mu\log t+O(\rcp t) \\
    &= \tau+\mu\log(\tau+\mu\log t+O(\rcp t))+O(\rcp t) \\
    &= \tau+\mu\log\tau+O(\frac{\log\tau}{\tau}).
  \end{split}
\end{equation}
Then, it is easy to confirm the following expressions.
\begin{equation}
  \label{eq:LRSVIIIasym}
  \begin{split}
  q_1(\tau) &= q_2(\tau)= (\tau+\mu\log \tau)^2+ O(\log\tau),  \\
  q_3(\tau) &= 4l^2\paren{1+\frac{2\mu}{\tau}}+O(\frac{\log\tau}{\tau^2}).
  \end{split}
\end{equation}
In particular, note that the fiber length
\begin{equation}
  \label{eq:LRSVIIIfiberlen}
  L\equiv\int_{\mathrm{fiber}}\sqrt{q_3}\s3
  =\int_0^{2\pi}\sqrt{q_3}dz= 2\pi\sqrt{q_3}
  =2\pi(2l+O(\rcp\tau))
\end{equation}
approaches constant $L_\infty\equiv\lim_{\tau\goes\infty}L=4\pi l$.
Therefore the parameter $l$ is $(4\pi)\inv$ times the
fiber length at the future infinity.

\section{Future asymptotics of scalar fields}
\label{sec:fa}

Let us consider the massless scalar field equation\footnote{In this
  section we employ the abstract index notation \cite{Wa} and use
  leading Latin letters $a,b,\cdots$ to denote abstract indices for
  vectors and tensors.}
\begin{equation}
  \label{eq:sceq}
  \ug ab\nabla_a\nabla_b \Psi=0,
\end{equation}
where $\nabla_a$ is the covariant derivative operator associated with
a spacetime metric $\dg ab$.

For our shift-free, Bianchi VIII metric \reff{eq:bmetric},
\begin{equation}
  \label{eq:BVIIIsc}
  \ug ab\nabla_a\nabla_b \Psi=
  \paren{\frac{-1}{\sqrt{-\ggg}}\vb t\paren{\sqrt{-\ggg}N^{-2}\vb t}
    +\Lap_q } \Psi,
\end{equation}
where $\sqrt{-\ggg}\equiv\sqrt{-\det\dg ab}=N\sqrt{q_1q_2q_3}$
and $\Lap_q$ is the Laplacian with respect to the spatial metric
$\dq ab$, which can be expressed using the invariant operators as
\begin{equation}
  \label{eq:BVIIILapq}
  \Lap_q =  \sum_{I=1}^{3}q_I\inv(\c I)^2.
\end{equation}

It is convenient to write the Laplacian $\Lap_q$ in terms of $\AA I$
and divide it into two parts, the \textit{homogeneous part}
$\Lap_q^{(0)}$ and the \textit{inhomogeneous part}
$\Lap_q^{(\mathrm{I})}$,
\begin{equation}
  \Lap_q= \Lap_q^{(0)}+\Lap_q^{(\mathrm{I})},
\end{equation}
where
\begin{equation}
  \begin{split}
    \Lap_q^{(0)} &= \rcp2(q_1\inv+q_2\inv)(\AA1\AA2+\AA2\AA1)
    -q_3\inv (\AA3)^2, \\
    \Lap_q^{(\mathrm{I})} &=
    \rcp2(q_1\inv-q_2\inv)((\AA1)^2+(\AA2)^2).
  \end{split}
\end{equation}
The ``homogeneous'' part does not change the index $m$ when it acts on
a single mode function $\phi_m=\phi_{m,\Lambda}$, while the
``inhomogeneous'' part does (and gives rise to inhomogeneous terms
in field equations). In fact, we find
\begin{equation}
  \label{eq:LapLapKm}
  \begin{split}
    \Lap_q^{(0)} \phi_m &= 
    \paren{\rcp2 (q_1\inv+q_2\inv)(a_{m-1}b_m+a_mb_{m+1})
    -q_3\inv m^2}\phi_m \\
  &= -K_m(t)\phi_m, \\
    \Lap_q^{(\mathrm{I})} \phi_m &=
    \rcp2(q_1\inv-q_2\inv)
    (a_m a_{m+1}\phi_{m+2}+b_mb_{m-1}\phi_{m-2}),
  \end{split}
\end{equation}
where
\begin{equation}
  K_m(t)\equiv \rcp2 (q_1\inv+q_2\inv)(m^2+\Lambda)+q_3\inv m^2.
\end{equation}
Here, we have used Eq.\reff{eq:b_m^2} to deform the first equation of
\reff{eq:LapLapKm}.

We expand the field component $\Psi=\Psi_\Lambda$ belonging to an
irreducible space specified by $\Lambda$ as
\begin{equation}
  \Psi(t,\x)=\sum_m \psi_m(t)\phi_m(\x).
\end{equation}
Then, the field equation \reff{eq:sceq} reduces to the following
equations:
\begin{equation}
  \label{eq:modeeqn}
  \frac{N}{(q_1q_2q_3)^{1/2}}\frac{d}{dt}
  \paren{\frac{(q_1q_2q_3)^{1/2}}{N}\frac{d\psi_m}{dt}}
  +N^2K_m(t)\psi_m
  = I_m(t),
\end{equation}
where
\begin{equation}
  I_m(t)\equiv
  \frac{N^2}{2}(q_1\inv-q_2\inv)
  (a_{m-2}a_{m-1}\psi_{m-2}+b_{m+2}b_{m+1}\psi_{m+2}).
\end{equation}
These equations form two sets of infinitely many simultaneous ODEs ---
one for $\{\psi_m\}_{m=\text{odd}}$ and one for
$\{\psi_m\}_{m=\text{even}}$.  The term $I_m$, which contains
$\psi_{m\pm2}$, works as an inhomogeneous term if we view the above
single equation as a dynamical equation for $\psi_m$.

\textbf{The LRS case}.
Let us consider the LRS background case \reff{eq:LRSsol}, in which,
due to the vanishing of the inhomogeneous term, each mode equation
\reff{eq:modeeqn} becomes independent. The equation becomes of the
form
\begin{equation}
  \label{eq:LRSeqn}
  \ddot\psi_m+\frac{\dot f}{f}\dot\psi_m+Z_m\psi_m=0,
\end{equation}
where
\begin{equation}
  Z_m(t)\equiv
  \frac{m^2}{4l^2}\paren{\frac{t^2+l^2}{f}}^2+\frac{m^2+\Lambda}{f},
\end{equation}
and
\begin{equation}
  f(t)\equiv t^2-l^2+2\mu t.
\end{equation}
Here, dot $(\,\dot{}\,)$ stands for $d/dt$.  We are interested in the
future ($t\goes +\infty$) asymptotic solution.  As emphasized in
\cite{Ta03}, to this it is necessary to transform the time variable to
one that is suitable for the analysis. Let us define the new time
variable $T$ by
\begin{equation}
  \frac{dT}{dt}=\sigma(t)>0,
\end{equation}
using a positive function $\sigma(t)$, which is to be determined.  We
transform the unknown function $\psi_m$ to another function $X_m$ so
that the new equation in terms of $X_m$, $dX_m/dT$, and $d^2X_m/dT^2$
has a vanishing $dX_m/dT$ term. It is easy to do this for given
$\sigma(t)$; The transformation is given by
\begin{equation}
  \label{eq:psialphaX}
  \psi_m=\alpha(t)X_m,
\end{equation}
where
\begin{equation}
  \alpha(t)\equiv \sigma^{-1/2}f^{-1/2}.
\end{equation}
The resulting equation is given by
\begin{equation}
  \label{eq:LRSTeqn}
  \frac{d^2X_m}{dT^2}+ W X_m=0,
\end{equation}
where
\begin{equation}
  W(T)\equiv \frac{1}{\sigma^2}
  \paren{\frac{\ddot\alpha}{\alpha}
    +\frac{\dot f}{f}\frac{\dot\alpha}{\alpha}+Z_m}.
\end{equation}
(Again, dot $(\,\dot{}\,)$ stands for $d/dt$, \textit{not} $d/dT$.)
If the function $W(T)$ in this equation approaches a constant $C\neq0$
as $T\goes \infty$ ($t\goes\infty$), then it may be natural to expect
that the equation \reff{eq:LRSTeqn} has fundamental solutions
approaching $e^{\pm i\sqrt{C}T}$ in case $C>0$, or $e^{\pm
  \sqrt{|C|}T}$ in case $C<0$. More precisely, using the standard
symbol $o(1)$ to signify a function such that $\lim_{t\goes\infty}
o(1)=0$, we expect that the equation has fundamental solutions of the
form $e^{\pm i\sqrt{C}T}(1+o(1))$ if $C>0$, or $e^{\pm
  \sqrt{|C|}T}(1+o(1))$ if $C<0$. Actually, this is the case if and
only if the function $W(t)$ satisfies the following finiteness
condition (e.g.,\cite{Ces});
\begin{equation}
  \label{eq:finiteness}
  \int^\infty\! |W-C|dT=\int^\infty\! |W-C|\sigma dt< \infty.
\end{equation}
In the present case (with $m\neq0$), it is confirmed that this
condition is satisfied with the choice $C=m^2/4l^2>0$ and
\begin{equation}
  \sigma(t)=1-\frac{2\mu}{t}.
\end{equation}
With this,
\begin{equation}
  T=\int \sigma(t)dt=t-2\mu\log t,
\end{equation}
and
\begin{equation}
  \alpha(t)= \frac{1}{t}+O(\rcp{t^3}).
\end{equation}
Recalling Eq.\reff{eq:psialphaX}, we thus have the following:
\begin{prop}
  \label{prop:LRSscgen}
  The generic, $m\neq0$, mode equation for massless scalar field on
  the LRS vacuum Bianchi VIII solution, satisfying
  Eq.\reff{eq:LRSeqn}, has the following fundamental solutions:
  \begin{equation}
    X_m^{(\pm)}(t)=
    {t\inv}e^{\pm i |\frac{m}{2l}|(t-2\mu\log t)}(1+o(1)).
  \end{equation}
\end{prop}

We must consider the $m=0$ case, separately. In this case, it is
possible to choose
\begin{equation}
  \sigma(t)= \frac{1}{t}
\end{equation}
so that
\begin{equation}
  W(t)= \Lambda-\rcp4 + O(\rcp{t}).
\end{equation}
Therefore, as long as $\Lambda\neq 1/4$, we can again apply the
criterion mentioned above, since
\begin{equation}
  |W-(\Lambda-\rcp4)|\,\sigma= O(\rcp{t^2}),
\end{equation}
implying the integral \reff{eq:finiteness} is finite with
$C=\Lambda-1/4\neq0$. Note that there are both possibilities of $C$
being positive and negative, depending on which we have two cases:
\begin{prop}
  \label{prop:LRSscU1}
  The $\U1$-symmetric, $m=0$, mode equation for massless scalar field
  on the LRS vacuum Bianchi VIII solution, satisfying
  Eq.\reff{eq:LRSeqn}, has the
  following fundamental solutions:
  \begin{equation}
    X_0^{(\pm)}(t)= 
    \begin{cases}
      t^{-\rcp2}e^{\pm i |s| \log t}(1+o(1)) & (\Lambda>\rcp4) \\
      t^{-\rcp2\pm |s|}(1+o(1)) & (0\leq\Lambda<\rcp4),
    \end{cases}
  \end{equation}
where $|s|\equiv \sqrt{|\Lambda-1/4|}$.
\end{prop}

We are interested in comparing with the Bianchi III model
\cite{TMY,Ta03}. To compare the asymptotic solutions, it is convenient
to use a proper time $\tau$ as canonical time. Since our time
coordinate $t$ asymptotically approaches proper time, it might be
expected that the formulas remain the same, but this is not exactly
the case. As confirmed in the following, the phase velocity of a part
suffers a slight modification.

Let $\tau$ be the proper time for the LRS solution, defined by
Eq.\reff{eq:LRSpropertime}. Using the formula
\reff{eq:LRSpropertimeinv}, it is easy to confirm, e.g.,
\begin{equation}
  t\inv=\tau\inv+O(\frac{\log\tau}{\tau^2}),
\end{equation}
and
\begin{equation}
  \begin{split}
    t-2\mu\log t &= \tau+\mu\log\tau+O(\frac{\log\tau}{\tau})
    -2\mu\log\tau(1+O(\frac{\log\tau}{\tau})) \\
    &= \tau-\mu\log\tau+O(\frac{\log\tau}{\tau}).
  \end{split}
\end{equation}
Beware that in the last equation the coefficient of $\log\tau$ is half
the one of $\log t$ in the left hand side.  Using these formulas, it
is easy to rewrite the asymptotic solutions in Propositions
\ref{prop:LRSscgen} and \ref{prop:LRSscU1} in terms of proper time
$\tau$. As a result, we have the following.
\begin{theorem}
  \label{th:LRSVIIIsc}
  A massless scalar field on a spatially compactified LRS vacuum
  Bianchi VIII solution can be decomposed into its mode components
  that are independent from each other and each of which is specified
  by the fiber eigenvalue $m$ and the Casimir eigenvalue $\Lambda$.
  Their fundamental solutions (at the future asymptotics) are given in
  terms of proper time $\tau$ as follows:
  \begin{equation}
    X_m^{(\pm)}(\tau)= 
    \begin{cases}
      {\tau\inv}e^{\pm i |\frac{m}{2l}|(\tau-\mu\log \tau)}(1+o(1))
      & (m\neq0), \\
      {\tau^{-\rcp2}}e^{\pm i |s|\log \tau}(1+o(1))
      & (m=0,\Lambda> 1/4) \\
      \tau^{-\rcp2\pm |s|}(1+o(1))
      & (m=0,0\leq\Lambda< 1/4)
    \end{cases}
  \end{equation}
  Here, $l$ and $\mu$ are the parameters in the LRS solution
  \reff{eq:LRSsol}, and $|s|\equiv \sqrt{|\Lambda-1/4|}$.
\end{theorem}

\section{Comparison with the Bianchi III model}
\label{sec:compari}

We make a comparison between the Bianchi VIII and Bianchi III systems.
We first compare the backgrounds, and then compare the asymptotic
behaviors of scalar field.

\textbf{The Bianchi III background solution.}
The LRS vacuum metric for Bianchi III is given
by
\begin{equation}
  \label{eq:LRSIIImetric}
  \ggg^{(\mathrm{III})}=-N^2(t)dt^2+q_1(t) ((\s1)^2+(\s2)^2)
  +q_3(t) (\s3)^2
\end{equation}
where
\begin{equation}
  \label{eq:LRSIIIsol}
  N^2(t)= \frac{t-\mu}{t+\mu}, \quad
  q_1(t)= (t-\mu)^2, \quad
  q_3(t)= 4l^2\frac{t+\mu}{t-\mu},
\end{equation}
and $\mu$ and $l$ are real parameters of solution. The invariant
1-forms $\s I$ ($I=1\sim 3$) here are those for Bianchi III and must
satisfy the following canonical relations;
\begin{equation}
  d\s1=\s1\^\s2,\quad
  d\s2=0,\quad
  d\s3=0.
\end{equation}
Our convention in terms of coordinates is
\begin{equation}
  \s1=dx/y,\quad
  \s2=dy/y,\quad
  \s3=dz.
\end{equation}
The invariant (dual) vectors are given by
\begin{equation}
  \c1=y\,\del/\del x,\quad
  \c2=y\,\del/\del y,\quad
  \c3=\del/\del z.
\end{equation}
In this section vectors $\xi_I$, $\c I$, one-forms $\s I$, and
metric functions $N(t)$, $q_I(t)$ all refer to those for Bianchi III
(not for Bianchi VIII), unless otherwise stated.

See Appendix A, \cite{TMY}, for details of the compactification of
this solution. We assume an \textit{orthogonal} \cite{TMY}
compactification, i.e., \textit{each} hyperbolic $z=\text{constant}$
surface is compactified to a higher genus surface $\Sigma_g$. Although
we do not repeat the details of the compactification, one of the most
important facts for our discussion is that when compactified, each
$z$-axis descends to circle fibers, and therefore the invariant vector
$\c3$ is called the \textit{fiber generator}, like in the Bianchi VIII
case.

As easily confirmed from the above metric, the conformal
base metric $h^{\mathrm{(III)}}= (\s1)^2+(\s2)^2$ is the same as the
one for the Bianchi VIII; they both coincide with the standard
hyperbolic metric $h\equiv (dx^2+dy^2)/y^2$. Since the 1-form $\s3$ is
dual to the fiber generator $\c3$, $q_3(t)$ is the scale factor for
the fibers, while $q_1(t)$ is the one for the base.  This
interpretation of the metric functions $q_I(t)$ is the same as that
for the LRS Bianchi VIII solution \reff{eq:LRSsol}.

The above metric is essentially the same one adopted in \cite{TMY}.
However, we have made two alterations. One is that we have replaced
the parameter $k$ with $-\mu$. As we will see, this makes the
correspondence to the Bianchi VIII metric \reff{eq:LRSsol} better. The
other alteration is the introduction of a new parameter $l$. At first,
this parameter might seem redundant, since we could set
$4l^2(\s3)^2\goes (\s3)^2$ by the induced map of the scaling
diffeomorphism $z\goes z/(2l)$.  However, we should notice that this
diffeomorphism is \textit{not} well defined on the compactified
manifold. To see this, let us focus on the fiber submanifold
$(\R,q_3(\s3)^2)$. To compactify we use the action by the translation
$z\goes z+2\pi$. This map generates a group, $2\pi\Z$. The circle
fiber $\mathcal{F}$ can therefore be expressed as $2\pi\Z\backslash
(\R,q_3(\s3)^2)$. The translation however does not commute with the
scaling diffeomorphism, which means that the scaling is not well
defined on the compactified manifold.  The parameter $l$ is therefore
\textit{not} redundant for the compactified manifold. The significance
of this parameter is apparent if we consider the length of the fiber,
which is given by
\begin{equation}
  \label{LIII}
  L^\mathrm{(III)}=\int_\text{fiber}\sqrt{q_3}\s3
  =\int_0^{2\pi}\sqrt{q_3}dz
  = 4\pi l\sqrt{\frac{t+\mu}{t-\mu}}.
\end{equation}
In particular,
$L^\mathrm{(III)}_\infty\equiv\lim_{t\goes\infty}L^\mathrm{(III)}=
4\pi l$. Therefore $l$ is the $(4\pi)\inv$ times the fiber length at
the future infinity, like the LRS Bianchi VIII case. (See
Eq.\reff{eq:LRSVIIIfiberlen}.)

We comment that we could also use the fiber metric $q_3'(\s3)^2$ that
is obtained by setting $4l^2=1$ as the universal cover metric, to
express the compactified fiber. To this, we need to allow the covering
group (rather than the metric) to have the parameter $l$, i.e., to
compactify the fiber we consider the one-parameter translation $z\goes
z+4\pi l$.  Then, it is easy to see that the resulting fiber
$\mathcal{F}'=4\pi l\Z\backslash (\R,q_3'(\s3)^2)$ is equivalent to
the fiber $\mathcal{F}$. For example, the fiber length is computed as
\begin{equation}
  \int_\text{fiber}\sqrt{q_3'}\s3
  =\int_0^{4\pi l}\sqrt{q_3'}dz
  = 4\pi l\sqrt{\frac{t+\mu}{t-\mu}},
\end{equation}
which agrees with $L^\mathrm{(III)}$. Remark that the parameter $l$ is
a relevant parameter in any case. In general, given a compactified
locally homogeneous manifold, there are two ways to express it
depending upon whether we fix the covering group or not.  See
\cite{TKH2} for a treatment of spatially compactified models with
fixed covering groups (i.e., with no dynamical degrees of freedom in
the covering group acting on spacetime), and \cite{TKH1} for one
with varying covering groups with the smallest number of parameters
in the universal cover metric. (See also \cite{AS,KTH,Ko} for related
discussions.) Note that in \cite{TMY} the background spacetime is
expressed in the latter view point, while in this paper the former
point of view has been exploited.

The reason we introduce the two-parameter metric \reff{eq:LRSIIIsol}
is that it has a straightforward correspondence to the (LRS) Bianchi
VIII metric \reff{eq:LRSsol}. In particular, as mentioned, the
parameter $l$ has the same meaning that it is $(4\pi)\inv$ times the
fiber length at the future infinity.

Let $\tau$ be the proper time defined by
\begin{equation}
  \begin{split}
    \tau &= \int N(t)dt=\int\sqrt{\frac{t-\mu}{t+\mu}}dt
    =\int \paren{1-\frac{\mu}{t}+O(\rcp{t^2})}dt \\
    &= t-\mu\log t+O(\rcp t).
  \end{split}
\end{equation}
This implies
\begin{equation}
  \label{eq:t->tauIII}
  t= \tau+\mu \log \tau +O(\frac{\log\tau}{\tau}).
\end{equation}
Therefore, from Eqs.\reff{eq:LRSIIIsol} we have
\begin{equation}
  \label{eq:LRSIIIasym}
  \begin{split}
    q_1(\tau) &= (\tau+\mu\log \tau)^2+O(\log\tau), \\
    q_3(\tau) &=
    4l^2\paren{1+\frac{2\mu}{\tau}}+O(\frac{\log\tau}{\tau^2}).
  \end{split}
\end{equation}
Comparing these equations with Eqs.\reff{eq:LRSVIIIasym} it is
confirmed that the asymptotic behavior of the LRS Bianchi III model is
the same as that of the Bianchi VIII at least up to second leading
terms. That is, the future asymptotic behaviors of base surface and
circle fiber are the same for both LRS Bianchi III and VIII solutions.
This establishes the \textit{background degeneracy} of the two models.

\textbf{Asymptotics of scalar fields on the Bianchi III.}  To compare
the behaviors of scalar field we need to summarize the asymptotic
properties of scalar field on the LRS Bianchi III background.  As far
as the compactified spatial manifold is of the orthogonal type, the
harmonics can be constructed by simply making products of those
$c_m(z)$ on the fiber and those $\hat S_\lambda(x,y)$ on the base
hyperbolic surface. Those harmonics are defined by the following
eigenvalue equations
\begin{equation}
  \label{eq:eigenIIIeqns}
  \begin{split}
    \c3 c_m &= im c_m, \\
    \hat\Lap_0 \hat S_\lambda &= -\lambda^2 \hat S_\lambda.
  \end{split}
\end{equation}
Here, $\hat\Lap_0$ is the Laplacian with respect to the standard
hyperbolic metric $h= (\s1)^2+(\s2)^2$. The harmonics (mode functions)
on an orthogonal closed Bianchi III manifold are the products
\begin{equation}
  S_{\lambda,m}=c_m \hat S_\lambda.
\end{equation}
We call $m$ the \textit{fiber eigenvalue}, and $\lambda^2$ the
\textit{base eigenvalue}. Remember that since we take the
two-parameter metric \reff{eq:LRSIIIsol}, the identifications along
the fibers are taken with the fixed step $2\pi$, and as a result we
have
\begin{equation}
  \label{eq:minZIII}
  m\in \Z,
\end{equation}
since the solution of the first equation in Eqs.\reff{eq:eigenIIIeqns}
is given by $e^{imz}$.

The mode-decomposed massless scalar field equation, computed with the
one-parameter metric, is given in Eq.(243), \cite{TMY}. To convert
this equation to the one computed with the two-parameter metric
\reff{eq:LRSIIIsol}, it is enough to perform the replacement
\begin{equation}
  m\goes \frac{m}{2l}.
\end{equation}
Then, the equation reads
\begin{equation}
  \label{eq:LRSIIIsceq}
  \ddot\psi+\frac{2t}{t^2-\mu^2}\dot\psi
  +\paren{\frac{\lambda^2}{t^2-\mu^2}
    +\paren{\frac{m}{2l}}^2\frac{(t-\mu)^2}{(t+\mu)^2}}\psi=0.
\end{equation}
The unknown function $\psi=\psi_{\lambda,m}(t)$ is the field component
for the mode $S_{\lambda,m}$. The exact solutions for the $\mu=0$
background are given in Appendix B, \cite{TMY}. No results are however
presented there for general $\mu$. Fortunately, however, it is not
difficult to obtain asymptotic solutions for the equation, following
the same procedure shown in the previous section. We just present the
result here.
\begin{prop}
  Consider the massless scalar field equation on the LRS Bianchi III
  vacuum metric \reff{eq:LRSIIImetric}, given by
  Eq.\reff{eq:LRSIIIsceq}.  The fundamental solutions of the equation
  are given by
\begin{equation}
  \label{eq:Ypmt}
  Y^{(\pm)}_m(t) =
  \begin{cases}
    t\inv e^{\pm i \abs{\frac{m}{2l}}(t-2\mu\log t)}(1+o(1))
    & (m\neq 0) \\
    t^{-\rcp2} e^{\pm i |s| \log t}(1+o(1)) 
    & (m= 0, \lambda^2>1/4) \\
    t^{-\rcp2 \pm |s|}(1+o(1))
    & (m= 0, 0\leq\lambda^2<1/4),
  \end{cases}
\end{equation}
where
\begin{equation}
  |s|\equiv \sqrt{|\lambda^2-\rcp4|}.
\end{equation}
\end{prop}

It is easy to convert $t$ into the proper time $\tau$ in this
solution. Using Eq.\reff{eq:t->tauIII}, we obtain the following.
\begin{theorem}
  A massless scalar field on a spatially compactified LRS vacuum
  Bianchi III solution can be decomposed into its mode components that
  are independent from each other and each of which is specified by
  the fiber eigenvalue $m$ and the base Laplacian eigenvalue
  $\lambda^2$. Their fundamental solutions (at the future
  asymptotics) are given in terms of proper time $\tau$ as follows:
\begin{equation}
  \label{eq:Ypmtau}
  Y^{(\pm)}_m(\tau) =
  \begin{cases}
    \tau\inv e^{\pm i \abs{\frac{m}{2l}}(\tau-\mu\log \tau)}(1+o(1))
    & (m\neq 0) \\
    \tau^{-\rcp2} e^{\pm i |s| \log \tau}(1+o(1)) 
    & (m= 0, \lambda^2>1/4) \\
    \tau^{-\rcp2 \pm |s|}(1+o(1))
    & (m= 0, 0\leq\lambda^2<1/4),
  \end{cases}
\end{equation}
Here, $k$ and $l$ are the parameters in the LRS solution
\reff{eq:LRSIIIsol}, and $|s|\equiv \sqrt{|\lambda^2-1/4|}$.
\end{theorem}
Again, only the difference from the version expressed in terms of the
coordinate time $t$ is that the numerical factor of $-2\mu\log t$, in
the phase part of the solution for $m\neq0$, becomes half,
$-\mu\log\tau$.

\textbf{Comparison.}  Comparing the above theorem with Theorem
\ref{th:LRSVIIIsc}, we see that the asymptotic solutions in the
Bianchi VIII and Bianchi III models completely agree, provided the
correspondence $\lambda^2\leftrightarrow \Lambda$ is understood. We
call this agreement the \textit{scalar field asymptotic degeneracy} of
the two models.

In the following we give a short account how this degeneracy can be
understood. First, note that an arbitrary scalar field solution $\Psi$
can be decomposed into two parts, the $\U1$-symmetric part
$\Upsilon(t,x,y)$ and the rest part
$\Psi_\mathrm{(gen)}(t,x,y,z)\equiv \Psi-\Upsilon$;
\begin{equation}
  \Psi=\Psi_\mathrm{(gen)}+\Upsilon.
\end{equation}
(We do not have to mode-decompose each part here.) We first consider
the $\U1$-symmetric part $\Upsilon$. Remember that the $\U1$-symmetry
means the translation symmetry along the fibers. As mentioned, as long
as the background is LRS the background is also $\U1$-symmetric along
the fibers for both Bianchi types. Therefore we can consistently
contract the fibers and obtain a reduced scalar field system on a
$(2+1)$-dimensional spacetime. The 2-dimensional spatial manifold is
the higher genus surface $\Sigma_g$, and the spacetime manifold
becomes $\R\times\Sigma_g$ for both models. It is easy to write down
the field equation on this contracted manifold for the Bianchi VIII
system. From Eqs.\reff{eq:BVIIIsc}, \reff{eq:BVIIILapq} and
\reff{eq:c12c22}, we have
\begin{equation}
  \label{eq:LRSBVIIIsc}
  \ug ab\nabla_a\nabla_b \Upsilon=
  \paren{\frac{-1}{\sqrt{-\ggg}}\vb t\paren{\sqrt{-\ggg}N^{-2}\vb t}
    +q_1\inv \hat\Lap_0 } \Upsilon,
\end{equation}
where $\hat\Lap_0$ is the Laplacian \reff{eq:hatLap0} with respect to
the standard hyperbolic metric. Thus, the field equation on the
contracted manifold is given by
\begin{equation}
  \label{eq:LRSU1sc}
  \paren{
    -\frac{N}{q_1\sqrt{q_3}}\vb t \paren{\frac{q_1\sqrt{q_3}}{N}\vb t}
  +N^2q_1\inv\hat\Lap_0}\Upsilon=0.
\end{equation}
In this equation, the functions $q_I(t)$ and $N(t)$ should be thought
of as the functions of time given in Eq.\reff{eq:LRSsol}. The equation
for the Bianchi III is also obtained in the same way, and we find that
the above equation is exactly valid for this case, too, provided that
$q_I(t)$ and $N(t)$ are those for the LRS Bianchi III solution
\reff{eq:LRSIIIsol}. This equivalence explains the degeneracy for the
$\U1$-symmetric modes, since we have the background degeneracy, which
means that the coefficient functions in Eq.\reff{eq:LRSU1sc} show
asymptotically the same behavior for both cases.

Next, consider the rest part $\Psi_\mathrm{(gen)}$, which we call the
``generic part.'' Note that the spatial Laplacians for the two LRS
backgrounds are expressed as
\begin{equation}
  \label{eq:LapqVIIIandIII}
  \Lap_q=
  \begin{cases}
    q_1\inv((\c1)^2+(\c2)^2)+q_3\inv (\c3)^2 & \text{(VIII)} \\
    q_1\inv\hat\Lap_0+q_3\inv (\c3)^2. & \text{(III)}
  \end{cases}
\end{equation}
The key fact is that in bath cases, $q_3\inv$ dominates $q_1\inv$;
\footnote{It is noteworthy that the dominance of $q_3\inv$ also
  holds for the non-LRS Bianchi VIII background, i.e., as we can check
  using a result in \cite{Hans}, it holds $\lim q_1\inv/q_3\inv=\lim
  q_2\inv/q_3\inv= 0$.}
\begin{equation}
  \lim_{t\goes\infty}\frac{q_1\inv(t)}{q_3\inv(t)}=0.
\end{equation}
Since from the assumption we have $\c3\Psi_\mathrm{(gen)}\neq 0$, this
may suggest that the term in $q_3\inv$ will dominate the term in
$q_1\inv$ in both Laplacians. If this can be justified, we have the
simplification
\begin{equation}
  \label{eq:Lapqgoesc3}
  \Lap_q\goes
    q_3\inv (\c3)^2,  \quad \text{(VIII and III)}
\end{equation}
resulting in the same asymptotic equation for both systems again;
\begin{equation}
  \paren{
    -\frac{N}{q_1\sqrt{q_3}}\vb t \paren{\frac{q_1\sqrt{q_3}}{N}\vb t}
  +N^2q_3\inv\c3}\Psi_\mathrm{(gen)} = 0.
\end{equation}
This explains the degeneracy for the generic modes.

We call the last equation the \textit{fiber term dominated (FTD)
  equation} of the scalar field equation.\footnote{This equation can
  be considered as an analogy of the asymptotically velocity term
  dominated (AVTD) equations \cite{IM} of Einstein's equation valid
  for the opposite time direction.}  Beware that in contrast to the
$\U1$-symmetric equation \reff{eq:LRSU1sc}, the above equation
\textit{cannot} be justified as a symmetry reduction. That is,
although we dropped off $\c1$ and $\c2$-dependent terms, it is not to
impose the additional conditions $\c1\Psi=\c2\Psi=0$.  Indeed, if this
is the case for the Bianchi VIII we must have
$\c3\Psi=-[\c1,\c2]\Psi=0$, implying only spatially constant
configurations are allowed. Therefore $\Psi$ has no symmetry and in
particular depends on all spatial coordinates $\Psi=\Psi(t,x,y,z)$.

Note that although Eq.\reff{eq:Lapqgoesc3} looks natural, we need a
proof for its justification, which is what we have done with the mode
decomposed equations.

\section{Conclusions}
\label{sec:con}

We have separated the scalar field equation on a Bianchi VIII
background using the harmonics for the circle-fibered closed Bianchi
VIII manifold. The reduced wave equations form two sets of infinite
number of simultaneous ordinary differential equations (ODEs). This is
a result of the fact that irreducible representations of a noncompact
Lie group like the Bianchi VIII group are in general infinite
dimensional. When the background is LRS, however, each single wave
equation becomes closed itself due to the additional symmetry. We have
analyzed this closed equation and obtained future asymptotic
solutions, as in Theorem \ref{th:LRSVIIIsc}. In particular, for the
$m=0$ ($\U1$-symmetric) and $\Lambda<1/4$ (``small eigenvalue'') case
the solution is non-oscillatory, while the other cases are
oscillatory.  They are all decaying (except one of the two fundamental
solutions for the ``zero-mode'' with $m=\Lambda=0$).

We have seen that this result completely agree with that of the
Bianchi III model. We interpret this ``scalar field
asymptotic degeneracy'' as an evidence that other linear fields,
including electromagnetic fields and linear perturbations, on the
Bianchi VIII background also have the same asymptotic behaviors as
those on the Bianchi III. Since many results are already known for the
Bianchi III \cite{TMY,Ta03}, we can, based on them, conjecture
corresponding properties. Of our interest is that of the linear
perturbations. The asymptotic solution of the stability measures (a
kind of normalized gauge-invariant variables) for the Bianchi III,
given in Theorem 2.5, \cite{Ta03}, shows that the Bianchi III vacuum
solution is asymptotically unstable, meaning that the perturbed
spacetime asymptotically becomes more and more inhomogeneous. It is
therefore plausible that the (LRS) Bianchi VIII vacuum solution is
also asymptotically unstable in the same sense.

One however needs to be careful about the difference between the
properties of the vector or tensor harmonics applied to the two
models. For example, the Bianchi III tensor harmonics
are split into four kinds, the ``even'' ones, the ``odd'' ones, the
``harmonic'' ones, and the ``transverse-traceless (TT)'' ones
\cite{TMY}.  Accordingly, there are four kinds of (independent)
perturbations in the Bianchi III case. On the other hand, we will have
no such splitting in the Bianchi VIII case.  This difference comes
from whether the invariant frame $\{\s I,\c I\}$ is well defined on
the compactified manifold considered, since if it is well defined we
can construct all vector or tensor harmonics from the scalar harmonics
with the help of the frame. See \cite{Ta04a} for an explicit example,
where a Bianchi II case is treated. Because of this property, the
Bianchi VIII tensor harmonics will not have a splitting like the
Bianchi III one.

Because of this difference, the reduced Bianchi VIII perturbation
equations will have somewhat different properties. In particular,
while, in the Bianchi III case, the reduced perturbation equations are
given as independent second order ODEs (or independent systems of two
simultaneous first order ODEs), in the Bianchi VIII case they will be
given as independent systems of \textit{two} simultaneous second order
ODEs (or independent systems of \textit{four} simultaneous first order
ODEs).  (Here, we are assuming an LRS background for both Bianchi
types.) We can interpret this increase of variables in an independent
system of equations as a consequence of the coupling between an
``even'' and ``odd'' mode, like the Bianchi II case \cite{Ta04a}.

The result we should refer to to obtain a corresponding property about
the Bianchi VIII perturbations should be the one for the ``even''
perturbations of the Bianchi III solution. This is because they are
the dominant perturbations among the even and odd ones. It is apparent
that the Bianchi VIII perturbations will have nothing to do with the
``harmonic'' and ``TT'' ones of the Bianchi III solution, since those
perturbations are connected to the modes that cannot be produced from
the scalar harmonics. As a result, we can conclude that the
conjectured growth rate of the stability measure for the generic
Bianchi VIII perturbations is $O(\tau)$ (Cf. Theorem 2.5,
\cite{Ta03}), meaning it is unstable.

Finally, we comment on non-LRS cases. Remember that the degeneracy
between the Bianchi VIII and III systems concerns the LRS backgrounds.
Although, since a Bianchi III manifold cannot be compactified unless
it is LRS, we do not consider non-LRS Bianchi III system, the non-LRS
Bianchi VIII system may be of great interest itself. However, the
apparent difficulty is that we must solve infinite number of
simultaneous ODEs in this case.

We point out that this difficulty might not be so crucial. We have
seen that the scalar field asymptotic degeneracy (for the generic
modes) was a consequence of the fiber term dominated (FTD) behavior.
Since the dominance of the scale factor function $q_3\inv$ over
$q_1\inv$ and $q_2\inv$ continues to hold for the non-LRS background,
we can expect that an FTD behavior holds for this case, too. If this
is justifiable, it implies that each single mode becomes (virtually)
independent like in the LRS case. (This is what can be easily
confirmed by inspecting the non-LRS case equation.) Therefore the FTD
behavior may be the key to find asymptotic solutions of the non-LRS
case field equations.

\section*{Acknowledgment}

The author thanks Vincent Moncrief for helpful conversations at an
early stage of this work.

\end{document}